\documentclass[review,onefignum,onetabnum]{siamonline250211}

\usepackage{lipsum}
\usepackage{amsfonts}
\usepackage{graphicx}
\usepackage{algorithm}
\usepackage{algpseudocode}
\usepackage{amsmath}
\usepackage{amssymb}
\usepackage{xcolor}
\usepackage{bbm}
\usepackage{epstopdf}
\ifpdf
  \DeclareGraphicsExtensions{.eps,.pdf,.png,.jpg}
\else
  \DeclareGraphicsExtensions{.eps}
\fi

\usepackage{enumitem}
\setlist[enumerate]{leftmargin=.5in}
\setlist[itemize]{leftmargin=.5in}

\newsiamremark{remark}{Remark}
\newsiamremark{hypothesis}{Hypothesis}
\crefname{hypothesis}{Hypothesis}{Hypotheses}
\newsiamthm{claim}{Claim}
\newsiamremark{fact}{Fact}
\crefname{fact}{Fact}{Facts}

\headers{Wavelet-based Global Orientation and  Surface Reconstruction for Point Clouds}{}

\title{Wavelet-based Global Orientation and  Surface Reconstruction for Point Clouds}
\author{\parbox{\textwidth}{\centering Yueji Ma $^{1,2}$, Yanzun Meng$^{1,2}$, Dong Xiao$^{3,4}$, Zuoqiang Shi$^{2,5}$ \thanks{Corresponding author, author’s address: zqshi@tsinghua.edu.cn} and Bin Wang$^{3}$  \\$^1$ Department of Mathematical Sciences, Tsinghua University, Beijing, China\\
         $^2$ Yau Mathematical Sciences Center, Tsinghua University, Beijing, China\\
       $^3$ School of Software, Tsinghua University, Beijing, China\\
       $^4$ School of Mathematical Sciences, University of Science and Technology of China, Hefei\\
       $^5$ Yanqi Lake Beijing Institute of
Mathematical Sciences and Applications, Beijing, China}
}

\usepackage{amsopn}
\DeclareMathOperator{\diag}{diag}

\ifpdf
\hypersetup{
  pdftitle={Wavelet-based Global Orientation and  Surface Reconstruction for sparse point clouds},
  pdfauthor={Anonymous},
}
\fi

\usepackage{algpseudocode}
\usepackage{multirow}
\usepackage{microtype}
\usepackage{amsmath}
\usepackage{lineno}
\usepackage{url}
\usepackage{caption}
\usepackage{graphicx}
\usepackage{amssymb}
\usepackage{amsmath}
\usepackage{mathrsfs}
\usepackage{booktabs}
\usepackage{bm}
\usepackage{bbm} 
\usepackage{subfigure}

\def\d{\ensuremath{\mathrm{d}}}

\def\cos{\ensuremath{\mathrm{cos}}}
\def\sin{\ensuremath{\mathrm{sin}}}

\nolinenumbers
\begin{document}
\maketitle

\begin{abstract}

Unoriented surface reconstruction is an important task in computer graphics and has extensive applications. Based on the  compact support of wavelet and orthogonality properties, classic wavelet surface reconstruction achieves good and fast reconstruction. However, this method can only handle oriented points. Despite some improved attempts for unoriented points, such as iWSR, these methods perform poorly on sparse point clouds. To address these shortcomings, we propose a wavelet-based method to represent the mollified indicator function and complete both the orientation and surface reconstruction tasks. We use the modifying kernel function to smoothen out discontinuities on the surface, aligning with the continuity of the wavelet basis function. During the calculation of coefficient, we fully utilize the properties of the convolutional kernel function to shift the modifying computation onto wavelet basis to accelerate. In addition, we propose a novel method for constructing the divergence-free function field and using them to construct the additional homogeneous constraints to improve the effectiveness and stability. Extensive experiments demonstrate that our method achieves state-of-the-art performance in both orientation and reconstruction for sparse models. We align the matrix construction with the compact support property of wavelet basis functions to further accelerate our method, resulting in efficient performance on CPU. Our source codes will be released on GitHub.
\end{abstract}



\section{Introduction}
Unoriented point clouds are easier to obtain from the real world and have extensive representations. Hence, unoriented surface reconstruction is an important task in computer graphics and has many applications including medical treatment, artifact restoration, street view visualization.

In recent years, several implicit reconstruction methods for unoriented points have been proposed, such as GCNO \cite{2023GCNO}, iPSR \cite{hou2022iterative} and iWSR \cite{iwsr}. Although  these global methods still suffer from high running time, handling noisy or sparse models, these method provide us with inspiration for solving the reconstruction. GCNO \cite{2023GCNO}  proposes a smooth nonlinear objective function to characterize the requirements of an acceptable winding-number field and converts the problem into an unconstrained optimization problem.  Iterative Poisson surface reconstruction (iPSR) \cite{hou2022iterative} is an improvement on existing mature SPR \cite{kazhdan2013screened}  to eliminate dependence on normals. Based on wavelet basis functions and flipping strategies, iWSR \cite{iwsr} reconstructs the mesh from large-scale unoriented point clouds.

Although iWSR has the advantage of fast computing speed and the ability to handle large-scale point clouds, it performs poorly on sparse point clouds. Due to the limited information contained in sparse point clouds, the unsuited selection for step size often leads to incorrect orientation when applying flipping strategies. Furthermore, in sparse problems, the quality of a method's output holds greater significance than computational speed. Therefore, we want to address the shortcomings of wavelet theory and achieve high-quality orientation and reconstruction for sparse point clouds.
We represent the mollified indicator function based on wavelet basis functions, construct and solve the optimization problem about its coefficients, which finally eliminate the dependence on the normals, can be considered as ``\textit{Unoriented Wavelet-based  Surface Reconstruction}'' (UWSR).
In detail, we use the mollified kernel function in the field of partial differential equation (PDE) to handle the indicator function. This approach effectively smoothens out discontinuities on the surface, aligning with the continuity of the wavelet basis function.. We shift the modifying computation onto wavelet basis functions by leveraging the properties of convolutional kernels to compute coefficients.  During coefficient computation, we exploit the multi-scale structure and compact support property of wavelet basis functions. This allows us to accumulate contributions from input point clouds within their support regions, enabling sparse matrix computations. In addition, to address the issue of insufficient equation quantity, we propose the novel method to constructing the divergence-free function field and using them to establish the additional homogeneous constraints by divergence theorem.  This further improves the quality of orientation and reconstruction.

Our method also demonstrates robust orientation and reconstruction capabilities on various point clouds, including wire-frame, real-scanning, and noisy models. Our source codes will be released on GitHub. 


    
    
    

The paper is organized as follows. The related works of our method are introduced in \cref{sec:Related Work}. Our new algorithm for globally orientation and surface reconstruction is in \cref{sec:method}. The detailed experimental results are in \cref{sec:experiments}, and the limitations and conclusion follow in
\cref{sec:conclusions}.

\section{Related Work}
\label{sec:Related Work}
  For the extensive applications, surface reconstruction has been widely researched in the past decades. The vast majority of surface reconstruction methods based on implicit functions. According to the dependence on normal, we categorize these methods into two categories: oriented surface reconstruction and unoriented surface reconstruction.


\subsection{Surface Reconstruction for Oriented  Point Clouds}

The radial basis function methods treat each input point as the center of a radial basis function and seek the optimal weights of these functions to approximate the signed distance function (SDF) of the target surface.  \cite{923379} and \cite{walder2006implicit} propose to use compactly supported radial basis functions to deal with the issues in large-scale point clouds. To improve the running speed, \cite{10.1145/383259.383266} applies the fast multipole method (FMM) to reduce the cost of evaluating the interpolant at $M$ points from $O(MN)$ to $O(M + N \log N)$ operations. To further improve the outputs and utilize normal information, \cite{macedo2011hermite}, \cite{ijiri2013bilateral}, and \cite{liu2016closed} propose to use the Hermite RBFs to interpolate the positions of points and normals together. Robust and efficient reconstructions are shown on real data captured from a variety of scenes in these papers.  

In \cite{dey2005adaptive} and \cite{kolluri2008provably}, implicit moving least squares (MLS) is used to fit local algebraic shapes and minimize the total squared errors to nearby points and normals. These methods output a good approximation to the signed distance function of the target surface.  \cite{amenta2004domain} attempts to widen the domain without generating additional surfaces away from the correct surface. In addition, \cite{merry2013moving} implements a GPU accelerated reconstruction system that is able to process massive data sets.

In addition to methods based on approximation and fitting, many methods are inspired by the divergence theorem \cite{pfeffer1986divergence} and transforming the problem into calculating an integral on the target surface. RSM \cite{kazhdan2005reconstruction}, WSR \cite{manson2008streaming} and BWSR \cite{ren2018biorthogonal} respectively choose the Fourier basis, orthogonal wavelet basis, biorthogonal wavelet basis, and calculate the corresponding coefficient to represent the indicator function. RSM \cite{kazhdan2005reconstruction} has a breakneck running speed at that time by using fast Fourier transforms (FFT) algorithm. WSR \cite{manson2008streaming}  utilizes the orthogonal wavelet basis functions and its compact support property to achieve the linear space/running complexity and has the ability to handle multi-scale reconstruction. BWSR \cite{ren2018biorthogonal} further expands the basis function from orthogonal wavelet basis to biorthogonal wavelet basis and improves the output reconstructions. GR \cite{2019GR}  introduces the Gauss formula to surface reconstruction problems and achieves good results in both orientation and reconstruction.

Another popular choice is handling the reconstruction problem by constructing the spatial Poisson equation. This novel idea is first proposed in PR \cite{kazhdan2006poisson}. SSD \cite{calakli2011ssd}, SPR \cite{kazhdan2013screened},  and EPR \cite{kazhdan2020poisson} respectively  add different constraint conditions to optimization problem and fit the smoothed indicator function near the target surface,  resulting in enhanced performance. To handle the statistical queries, SPSR \cite{2022SPSR} introduces the stochastic Poisson surface reconstruction, a statistical derivation of PR as conditional probability distributions in a Gaussian Process. 

\subsection{Surface Reconstruction for Unoriented Point Clouds}
Most of point clouds from real-world only have position information and lacking the oriented normal. Though addressing this challenge is demanding, unoriented surface reconstruction has a wider
and potential applications with considerable attention.

\paragraph{\textbf{Reconstruction based on Deep Learning}}
With the powerful neural representation capabilities, deep learning methods have been intensively studied and widely used in surface reconstruction in recent years. These methods represent implicit fields with neural networks. Point2surf (P2S) \cite{erler2020points2surf} improves generalization performance and reconstruction accuracy by learning a prior over a combination of detailed local patches and coarse global information. Neural-Pull \cite{ma2021neural} trains a neural network to pull query 3D locations to their closest points on the surface using the predicted signed distance values and the gradient at the query locations.  DiGS \cite{ben2022digs} incorporates second-order derivative constraints to guide the implicit neural representation's (INR's) learning process, leading to better representations. Neural-Singular-Hessian (NSH)  \cite{wang2023neural} enforces the Hessian of the neural implicit function to have a zero determinant for points near the surface, which suppresses ghost geometry and recovers details from unoriented point clouds. Nevertheless, these deep learning methods tend to be extensive training and long running time.

\paragraph{\textbf{Reconstruction based on Propagation or Iteration}} An alternative strategy involves transforming unoriented point clouds into oriented ones and leveraging established surface reconstruction methods to accomplish the task. The key aspect of these methods is the propagation of orientation. 
The initial orientation propagation method dates back to 1992. \cite{hoppe1992surface} constructs a graph for input points, whose edge costs represented orientation consistency. After that, the minimum spanning tree (MST)  is employed to find the solution. While this straightforward approach lacks robustness, in recent years, many methods have been proposed to improve it. Dipole \cite{dipole} utilizes networks to predict consistently oriented normals for local patches, followed by dipole-based propagation to attain global consistency. IWSR \cite{iwsr} uses flipping-based iterative algorithms and proposes two novel criteria for flipping to solve the simplified problem. Iterative Poisson surface reconstruction (iPSR) \cite{hou2022iterative} calculates the normals from the surface in the preceding iteration and then generates a new surface with better quality. Nevertheless, these methods are highly sensitive to flipping criteria, graph weight selection, or the hyperparameters. Challenges persist in handling complex geometries effectively.

\paragraph{\textbf{Reconstruction based on Indicator Function}} A popular choice in this field is reconstructing the surface by representing the indictor function $\chi$ of a bounded area and making the input unoriented points located on the boundary of the region exactly. Based on GR \cite{2019GR}, parametric Gauss reconstruction (PGR) \cite{2022PGR} takes linearized surface elements as the unknowns and constructed the linear equations.It uses conjugate gradient method to solve such equation systems and calculate the indicator field. GCNO \cite{2023GCNO} proposes a new double well objective function to find the globally consistent normal orientations starting  from a set of random normals.

\section{Method}

\label{sec:method}

A critical and popular way to reconstruct the surfaces from the unoriented point clouds is by calculating the indicator function $\chi$ of target region. In detail,
let $\Omega \subset \mathbb{R}^{3}$ be an open and bounded region with the smooth boundary $\partial \Omega$ and the input oriented points are $\mathcal{P}=\{\bm{p}_i\}_{i=1}^{M}$ on the target surface $\partial \Omega$. Then,
\begin{equation}\label{12}
	\chi(\bm{x})=\begin{cases}
	1, & \mbox{if  } \bm{x} \in \Omega\\
	0,  & \mbox{if  } \bm{x} \notin \overline{\Omega},
\end{cases}
\end{equation}
where $\overline{\Omega}$ is the closure of $\Omega$, and we can use the indicator function to distinguish between inside and outside the region and find the position of surface.
 


\subsection{\textbf{Mollified Indicator Function}}
However, the indicator function $\chi(\bm{x})$  is discontinuous at the boundary of region and its value  is independent of the distance to the boundary, near the boundary. It greatly reduces the accuracy of generating surface meshes. Hence, we modify it to make it continue and contain the information of distance near the surface.  That is
\begin{align*}
    \overline{\chi}(\bm{x})=\int_{\mathbb{R}^3}\overline{K}_{\varepsilon}(\bm{x}-\bm{y})\chi(\bm{y}) \text{d} \bm{y},
    \end{align*}
    where: $\overline{K}_{\varepsilon}(\bm{x})$ is a mollifier function in three-dimension, and $\overline{\chi}(\bm{x})$ is the mollified indicator function and it can be continuous throughout the entire three-dimensional space. Taking any function $B(\bm{x})$ in three-dimensional space as the test function and employing the divergence theorem, its inner product with the mollified indicator function $\overline{\chi}(\bm{x})$ can be calculated as

    \begin{equation}\label{1}
	\begin{aligned}
		c_{i}=\langle \overline{\chi}(\bm{x}),B(\bm{x}) \rangle&=
		\int_{\mathbb{R}^{3}}\overline{\chi}(\bm{x})\cdot B(\bm{x}) \d \bm{x} =\int_{\mathbb{R}^3}\int_{\mathbb{R}^3}\overline{K}_{\varepsilon}(\bm{x}-\bm{y})\chi(\bm{y})\  \d \bm{y} \ B(\bm{x}) \d \bm{x} \\
        &=\int_{\mathbb{R}^3}\int_{\mathbb{R}^3}\overline{K}_{\varepsilon}(\bm{x}-\bm{y})\chi(\bm{y}) \ B(\bm{x}) \  \d \bm{y} \d \bm{x}\\
        &=\int_{\mathbb{R}^3}\int_{\mathbb{R}^3}\overline{K}_{\varepsilon}(\bm{y}-\bm{x})\chi(\bm{x}) \ B(\bm{y}) \  \d \bm{y} \d \bm{x}\\
        &=\int_{\mathbb{R}^3}\chi(\bm{x})\int_{\mathbb{R}^3}\overline{K}_{\varepsilon}(\bm{y}-\bm{x}) \ B(\bm{y}) \  \d \bm{y} \d \bm{x}\\
        &=\int_{\mathbb{R}^3}\chi(\bm{x})\overline{B}_{\varepsilon}(\bm{x}) \d \bm{x}=\int_{\Omega} \overline{B}_{\varepsilon}(\bm{x}) \d \bm{x} \\
        &=\int_{\partial \Omega}\bm{\overline{F}}_{\varepsilon}(\bm{x}) \cdot \bm{n}(\bm{x}) \d S,
	\end{aligned}
\end{equation}
where $\langle \cdot ,\cdot \rangle$ represents the $L_2$ inner product in $\mathbb{R}^3$ and  $\nabla \cdot \overline{F}_{\varepsilon}(\bm{x})=\overline{B}_{\varepsilon}(\bm{x}) $.
The mollifying has been transferred to the test function $B(\bm{x})$. We choose the Db4 wavelet bases as the test functions with the following reasons. (1) The Db4 wavelet bases exhibit orthogonality, with the corresponding theorems detailed in the Appendix \ref{Appendix}. (2) The Db4 wavelet bases has a compact support set of $7^3$ in $\mathbb{R}^3$. When $level=3$, its support becomes $(\frac{7}{2^3})^3$ fitting the normal size of octree ($[0,1]^3$) well, which can reduce computational complexity greatly.
(3) Db4 with $level=3$ is the wavelet basis function with the highest vanishing moment among the wavelet bases that can be accommodated in $[0,1]^3$. The larger the vanishing moment of a wavelet basis function, the better its smoothness.

Similar to iWSR \cite{iwsr},
we truncate the Db4's infinite term representation of the mollified indicator function to a finite term 
\begin{equation}\label{11}
	\overline{\chi}(\bm{x}) = \sum_{i=1}^{N}c_{i}B_{i}(\bm{x}).
\end{equation}
  
The computation of the implicit field ( $\overline{\chi} (\bm{x})$) is simplified by calculating the corresponding coefficients $c_i$ of the wavelet bases. For any fixed $i$, due to the orthogonality, 
\begin{align*}
    \langle \overline{\chi}(\bm{x}),B_i(\bm{x}) \rangle=\langle \sum_{j=1}^{N}c_{j}B_{j}(\bm{x}),B_i(\bm{x}) \rangle=\sum_{j=1}^{N}  c_{j}\langle B_{j}(\bm{x}),B_i(\bm{x}) \rangle=c_i,
\end{align*}
these coefficients can be calculated by Equation \eqref{1}. During the calculation of $c_i$, we need handle the mollifying of Db4 wavelet bases. While it is possible to mollify each basis function individually, the number of wavelet basis functions can reach millions as the number of layers increases. Consequently, this approach is impractical. To tackle this dilemma, we propose a novel mollifying strategy to accelerate the speed of computation significantly by constructing a tensor-product-based mollifying kernel function,

 \begin{align*}
    &\overline{K}_{\varepsilon}(x,y,z)=K_{\varepsilon}(x)K_{\varepsilon}(y)K_{\varepsilon}(z)\\
    &K_{\varepsilon}(x)=\dfrac{1}{\varepsilon}K(\dfrac{x}{\varepsilon}),
\end{align*}
where $\varepsilon$ is the chosen smooth width, and $K(x)$ is the mollifier function in field of partial differential equations \cite{evans2022partial} in mathematics.
\begin{equation*}
	K(x)=\begin{cases}
	\dfrac{1}{\int_{-1}^{1}\exp(\dfrac{1}{x^2-1})\text{d} x} \cdot \exp(\dfrac{1}{x^2-1}), & \mbox{if  } |x|<1\\
	0,  & \mbox{if  } |x|\geqslant 1.
\end{cases}
\end{equation*}

 We only need to mollify the one-dimensional wavelet basis functions by utilizing tensor products. Taking $B(\bm{x})=\psi^{(1,0,0)}_{j,\bm{k}}(\bm{x})=\psi_{j,k_1}(x_1)\varphi_{j,k_2}(x_2)\varphi_{j,k_3}(x_3)$ as an example  and the detailed calculation process for mollifying is shown below. 

\begin{align*}
&\overline{B}_{\varepsilon}(\bm{x})=\int_{\mathbb{R}^3}\overline{K}_{\varepsilon}(\bm{y}-\bm{x}) \ B(\bm{y}) \  \d \bm{y}\\
&=\int_{\mathbb{R}}\int_{\mathbb{R}}\int_{\mathbb{R}}K_{\varepsilon}(y_1-x_1)K_{\varepsilon}(y_2-x_2)K_{\varepsilon}(y_3-x_3)\psi_{j,k_1}(y_1)\psi_{j,k_2}(y_2)\varphi_{j,k_3}(y_3) \d y_1 \d y_2 \d y_3\\
&= \int_{\mathbb{R}}K_{\varepsilon}(y_1-x_1)\psi_{j,k_1}(y_1) \d y_1 \cdot\int_{\mathbb{R}}K_{\varepsilon}(y_2-x_2)\varphi_{j,k_2}(y_2) \d y_2 \cdot\int_{\mathbb{R}}K_{\varepsilon}(y_3-x_3)\varphi_{j,k_3}(y_3) \d y_3 \\
&= \overline{\psi}_{\varepsilon,j,k_1}(x_1) \cdot \overline{\psi}_{\varepsilon,j,k_2} (x_2)\cdot\overline{\varphi}_{\varepsilon,j,k_3}(x_3).
\end{align*}

By utilizing the multi-scale property of wavelet basis functions, we can further transform all the mollifying of different layer basis functions into scale basis functions or wavelet basis functions to accelerate. Taking the wavelet basis function with translation $k$ in level $j$ ($\psi_{j,k}$) as the example,
\begin{equation}
\begin{aligned}
\overline{\psi}_{\varepsilon,j,k}&=\int_{\mathbb{R}}K_{\varepsilon }(y)\psi_{j,k}(x-y) \d y \\
&=\int_{\mathbb{R}}K_{\varepsilon }(y) \cdot 2^{\frac{j}{2}} \cdot \psi(2^j(x-y)-k)\d y \\
&=\int_{\mathbb{R}}K_{2^j \varepsilon }(2^j y)\cdot 2^j \cdot 2^{\frac{j}{2}} \cdot  \psi(2^j x- 2^j y-k)\d y\\
&= \int_{\mathbb{R}}K_{2^j \varepsilon}(t)\cdot 2^j \cdot 2^{\frac{j}{2}} \cdot  \psi(2^j x- t -k) \cdot 2^{-j} \d t\\
&=\int_{\mathbb{R}}K_{2^j \varepsilon}(t) \cdot 2^{\frac{j}{2}} \cdot  \psi(2^j x- t -k) \d t\\
&=2^{\frac{j}{2}}\overline{\psi}_{2^j \varepsilon}(2^j x-k),
\end{aligned}
\end{equation}
 it can also be obtained by processing the standard wavelet basis function $\psi$ using mollifier function with a width of $2^j \varepsilon$.
 
 The corresponding vector function $\overline{\bm{F}}_{\varepsilon}(\bm{x})$ can be constructed by tensor products, taking $\bm{e}=(1,0,0)$ as the example.

\begin{align}\label{12}
\overline{\bm{F}}_{\varepsilon,j,\bm{k}}^{(1,0,0)}(x_1,x_2,x_3)=(\overline{\Psi}_{\varepsilon,j,k_1}(x_1)\overline{\varphi}_{\varepsilon,j,k_2}(x_{2})\overline{\varphi}_{\varepsilon,j,k_3}(x_3),0,0),
\end{align} 
where $\overline{\Psi}_{\varepsilon}(t)=\int_{-\infty}^{t}\overline{\psi}_{\varepsilon}(s)\d s$, and
\begin{align*}
    \overline{\Psi}_{\varepsilon,j,k}(x)&=\int_{0}^{x}\overline{\psi}_{\varepsilon,j,k_1}(t) \d t\\
    &= \int_{0}^{x} 2^{\frac{j}{2}} \cdot\overline{\psi}_{2^j \varepsilon}(2^j t-k) \d t\\
    &=\int_{0}^{2^j x-k} 2^{\frac{j}{2}} \cdot \overline{\psi}_{2^j \varepsilon}(s)\cdot 2^{-j} \d s\\
    &=2^{-\frac{j}{2}} \overline{\Psi}_{2^j \varepsilon}(2^j x-k).
\end{align*}

Thus, we have achieved rapid computation of the mollified indicator function. We select the input points $\bm{p}_{i}$ lying on the surface as the query point $\bm{x}$ in Equation \eqref{chi} sequentially. According to  $\chi(\bm{p}_{i})=\frac{1}{2},\ i=1,2,\cdots,M$, We can construct $M$ non-homogeneous equations about the target surface.

\subsection{\textbf{The Homogeneous Constraint}}
Although we have constructed a series of non-homogeneous linear equations using the divergence theorem, only $M$
non-homogeneous equations are not enough, which is also shown in PGR \cite{2022PGR}. Since there are $3M$ unknowns, in order to avoid solving the underdetermined equation systems, we extensively utilize the well-known divergence theorem in vector analysis to construct additional homogeneous constraints.

For any  function $G: \mathbb{R}^3 \to \mathbb{R}^3$ with the continuous two-order partial derivatives, we denote $G=(G_x(x,y,z),G_y(x,y,z),G_z(x,y,z))$. Then, its curl field

\begin{align}\label{F}
    F=\nabla \times G=(\frac{\partial G_{z}}{\partial y}-\frac{\partial G_{y}}{\partial z},\frac{\partial G_{x}}{\partial z}-\frac{\partial G_{z}}{\partial x}
    ,\frac{\partial G_{y}}{\partial x}-\frac{\partial G_{x}}{\partial y})
\end{align} is the divergence-free vector field in three-dimension. In other words, 

\begin{align*}
    \text{div} F=\frac{\partial^{2} G_{z}}{\partial y\partial x}-\frac{\partial^{2} G_{y}}{\partial z\partial x}+\frac{\partial^{2} G_{x}}{\partial z\partial y}-\frac{\partial^{2} G_{z}}{\partial x\partial y}+\frac{\partial^{2} G_{y}}{\partial x \partial z}-\frac{\partial^{2} G_{x}}{\partial y \partial z}=0,
\end{align*}
and  
\begin{align}\label{h}
    \int_{\partial \Omega} F \cdot n \d S=\int_{\Omega} \text{div} F \d \bm{x}= \int_{\Omega} 0 \d \bm{x}=0.
\end{align}

Thus, this naturally construct more constraints on the surface. Since the target surface is closed and has no boundary, the same conclusion can be derived by the Stokes' theorem, which is shown in \cite{gotsman2024linear}. Equation \eqref{F} shows a method to construct the divergence-free function, and  any number of such homogeneous constraint can be constructed by selecting different kernel functions $G$. 

Since the linearly independent kernel functions can add the new information, we show the in-depth discussion on the selection of kernel functions $G$. We propose a novel way to constrcuting unlimited number of linearly independent functions via the  parameter $\bm{w_1}=(w_{1,1},w_{1,2},w_{1,3}) \in \mathbb{R}^3$ and $\bm{w_2}=(w_{2,1},w_{2,2},w_{2,3}) \in \mathbb{R}^3$, which are the weights of the points and functions, respectively. In detail, for any points $\bm{x}=(x_1,x_2,x_3)$, 
\begin{equation}\label{380}
\begin{aligned}
    &G_x= w_{2,1} \cdot \cos(w_{1,1}x_1+w_{1,2}x_2+w_{1,3}x_3),\\
    &G_y= w_{2,2} \cdot \sin(w_{1,1}x_1+w_{1,2}x_2+w_{1,3}x_3),\\
    &G_z= w_{2,3} \cdot \sqrt{w_{1,1}x_1+w_{1,2}x_2+w_{1,3}x_3}.\\
\end{aligned}  
\end{equation}

Although the proof that $G_x,G_y,G_z$ is a linearly independent set of functions for all $\bm{w_1}$ is difficult and complex due to their multivariate nature, it could be obvious in the one-dimension case. 
For the three-dimensional function family $G(\bm{w}_1)$, if one dimension of the function family is linearly independent (e.g. $G_x(\bm{w}_1)$) for the parameters, then $G(\bm{w}_1)$ must be linearly independent.

Actually, for $\cos(ax)$,  if $\sum_{k=1}^n c_k \cdot \cos(a_k x) = 0 \  (\forall x \in \mathbb{R})$, we take  Fourier transform on both sides of the equation in the sense of generalized function. Then,
\begin{align*}
    \mathcal{F}\left\{ \sum_{k=1}^n c_k \cdot \cos(a_k x) \right\} = \sum_{k=1}^n c_k \cdot \mathcal{F}\{\cos(a_k x)\}=0,
\end{align*}
where $\mathcal{F}$ represents the Fourier transform of function. Since
\begin{align*}
    \mathcal{F}\{\cos(a_k x)\} = \pi \left[ \delta(\omega - a_k) + \delta(\omega + a_k) \right],
\end{align*}
then,
\begin{align*}
    \sum_{k=1}^n c_k \pi \left[ \delta(\omega - a_k) + \delta(\omega + a_k) \right] = 0.
\end{align*}

The $\delta (\omega-a_k) $ corresponding to different $a_k $ do not overlap with each other in the frequency domain. In other words, in generalized function theory, $\{\delta (\omega - a_k) \}_{k=1}^{n}$ is linearly independent for any $a_k$. Hence, $\forall k$
\begin{align*}
    c_k \pi \left[ \delta(\omega - a_k) + \delta(\omega + a_k) \right] = 0. \quad 
\end{align*}

From this, it can be inferred that $c_k=0$ holds for all $k $. Hence, $G_x$ is a linearly independent set of functions about $\bm{w_1}$.   

In addition, in Equation \eqref{380}, $\bm{w_2}$ serves as a linear parameter. Its different values are selected mainly to enhance the numerical stability of the method.

We have provided an efficient method for constructing linearly independent kernel functions, but the way to generate such kernel functions is not unique. In practical applications, these homogeneous constraints can also be constructed through other different ways. For example, another way  is using the center of the function $\bm{c}=(c_1,c_2,c_3)$, let 
\begin{align*}
&G^{\bm{c}}=\left(\dfrac{1}{\bm{r}},\dfrac{1}{\bm{r}},\dfrac{1}{\bm{r}}\right),\\
    &r(x,y,z)=||(x,y,z)-(c_1,c_2,c_3)||,
\end{align*}
where $||\cdot||$ represents the $L_2$ norm in $\mathbb{R}^3$. The detail process of construction can refer to \cite{gotsman2024linear}.

\subsection{\textbf{Discretization and Constructing Linear System}}
After constructing the constraints, we then  discretize them to the linear system. From the Equation \eqref{1},
\begin{align*}
    c_i=\int_{\partial \Omega}\bm{\overline{F}}_{i;\varepsilon}(\bm{x}) \cdot \bm{n}(\bm{x}) \d S \approx \sum_{j=1}^{M}\bm{\overline{F}}_{i;\varepsilon}(\bm{p}_j) \cdot \bm{n}_{\bm{p}_j} \sigma_{\bm{p}_j}.
\end{align*}

 In the sum, $\bm{\overline{F}}_{i;\varepsilon}(\bm{p}_i) $ are known variables, and \textbf{the only unknowns are the normal $\bm{n}_{\bm{p}_{j}}$ and area elements $\sigma_{\bm{p}_{j}}$}, which is similar to \cite{2022PGR}.  We denote the $3$-dimensional vector
\begin{equation}\label{LSE}
    \bm{\mu}_{j}=(\mu_{j,1},\mu_{j,2},\mu_{j,3})  \triangleq \bm{n}_{\bm{p}_{j}} \sigma_{\bm{p}_{j}}
\end{equation}
to estimate area and normal information of $ \bm{p}_{j}$, called linearized surface element (LSE). Then, the calculation of coefficients $c_i$ has a discrete form $A\bm{\mu}$ and the indicator function can be represented as
\begin{align}\label{chi}
    \overline{\chi}(\bm{x})=\sum_{j,k}\psi^{\bm{e}}_{j,k}(\bm{x}) \cdot \left(\sum_{i=1}^{M}\bm{\overline{F}}^{\bm{e}}_{\varepsilon,j,k}(\bm{x}_i)\cdot\bm{\mu}_i \right).
\end{align}

Although the lack of outward normals prevents a direct estimation of $\bm{\mu}_{i}$, we can take them as unknown variables. Therefore, for an input point cloud of $M$ points, the number of unknown parameters is $3M$.

Firstly, the property that the indicator function is $\frac{1}{2}$ at input points can be discretized to construct the non-homogeneous constraints. We denote it as
\begin{align*}
    B_{f}(\mathcal{P}) A(\bm{p}_i;\mathcal{P})\bm{\mu}_i=\dfrac{1}{2}, \ i=1,2,\cdots M.
\end{align*}

In response to the insufficient quantity, we construct more homogeneous constraints based on the previous analysis in Equation \eqref{h}. In detail, we randomly select $N_{h}$ different sets of parameters $\bm{w_1}=(w_{1,1},w_{1,2},w_{1,3}) \in \mathbb{R}^3$,  $\bm{w_2}=(w_{2,1},w_{2,2},w_{2,3}) \in \mathbb{R}^3$ and take $\bm{\mu_{i}}$ as unknown variables and the constraints can be represented as 
\begin{align*}
    H_{G}(\bm{w}_{1}^{j},\bm{w}_{2}^{j}) \bm{\mu}_i=0, \ i=1,2,\cdots M ,\ j=1,2,\cdots N_{h}.
\end{align*}

By combining the above constraints, we can obtain the final equation system
\begin{align}\label{BA}
    \begin{bmatrix}
        B_{f}A\\
         H_{G}
    \end{bmatrix}\bm{\mu}=\begin{bmatrix}
                    \bm{b}_1\\
                    \bm{0}
                \end{bmatrix},
\end{align}
where $\bm{b}_1$=$\dfrac{1}{2}\in \mathbb{R}^{M\times 1}$
and denote it as $B\bm{\mu}=\bm{b}$ for brevity.

\subsection{\textbf{Solving the Linear System}}
After determining the strategy for constructing equations, the following subsection explains how to obtain the point cloud’s normal and complete the reconstruction task.

\paragraph{$N_{h} < 2 M$}
Under such circumstances, the matrix $B$  is not a square and the system \eqref{BA} is under-determined. The solution is not unique. Then, we try to find the minimal-norm solution:

\begin{align}\label{min}
\mathop{ \min }\limits_{\bm{\mu}} ||\bm{\mu}||_{2}^2 \quad  \text{subject to } B\bm{\mu}=\bm{b}.
\end{align}

In addition, $B$ may be ill-conditioned and lead to a numerically unstable solution. Since the solution of optimization problem  \eqref{min} is $B^T(BB^T)^{-1}\bm{b}$, we first assume $\bm{\mu}$ has the form $\bm{\mu}=B^T\bm{\xi}$. Then, the  optimization problem  \eqref{min} can be equivalently written as

\begin{align}\label{min2}
\mathop{ \min }\limits_{\bm{\mu}',\bm{\xi}} ||\bm{\mu}'-B^T\bm{\xi}||_{2}^2 \quad  \text{subject to } B\bm{\mu}'=\bm{b},\bm{\mu}=B^T\bm{\xi}.
\end{align}

To avoid the numerical instability mentioned above and to alleviate the possible interference of noise, we
apply regularization to the intermediate variable $\bm{\xi}$ based on the formulation \eqref{min2}:

\begin{align}\label{min3}
\mathop{ \min }\limits_{\bm{\mu}',\bm{\xi}} ||\bm{\mu}'-B^T\bm{\xi}||_{2}^2 + +||\Gamma \bm{\xi}||_{2}^{2}  \quad  \text{subject to } B\bm{\mu}'=\bm{b},\bm{\mu}=B^T\bm{\xi},
\end{align}
where $\Gamma$ is a matrix with $M$ columns, intended for regularization.

\begin{theorem}[\textbf{minimal-norm solution}]\label{TH3}
	The solution for optimization problem \eqref{min3} needs to satisfy $\bm{\xi}=(BB^T+\Gamma^T\Gamma)^{-1}\bm{b}$, and $\bm{\mu}=B^T\bm{\xi}$.
\end{theorem}
  \begin{proof}
  The equivalence can be  proved by the Lagrange multiplier method (LMM) \cite{bertsekas2014constrained}, and let 
  \begin{align*}
      f(\bm{\lambda},\bm{\xi},\bm{\mu})=\dfrac{1}{2}(||\bm{\mu}-B^T\bm{\xi}||^2_2+||\Gamma\bm{\xi}||_{2}^{2})+\bm{\lambda}^T(B\bm{\mu}-\bm{b}),
  \end{align*}
  then the KKT condition of optimization problem \eqref{min3} is
  \begin{equation}
  \begin{cases}
      &\dfrac{\partial f}{\partial \bm{\lambda}}=B\bm{\mu}'-\bm{b}=0\\
       &\dfrac{\partial f}{\partial \bm{\xi}}=BB^T\bm{\xi}-B\bm{\mu}'+\Gamma^T\Gamma\bm{\xi}\\
        &\dfrac{\partial f}{\partial \bm{\mu}'}=\bm{\mu}'-B^T\bm{\xi}+B^T\bm{\lambda}.\\
    \end{cases}
  \end{equation}

  From the first and second lines, we can get $(BB^T+\Gamma^T\Gamma)\bm{\xi}=\bm{b}$. Hence, $\bm{\xi}=(BB^T+\Gamma^T\Gamma)^{-1}\bm{b}$. 
  \end{proof}

  Thus, we construct the equations and can solve it using the conjugate gradient method \cite{nazareth2009conjugate} to avoid the calculations of matrix inversion. 

\begin{figure}[htbp]
    \centering
    \includegraphics[width=1.0\linewidth]{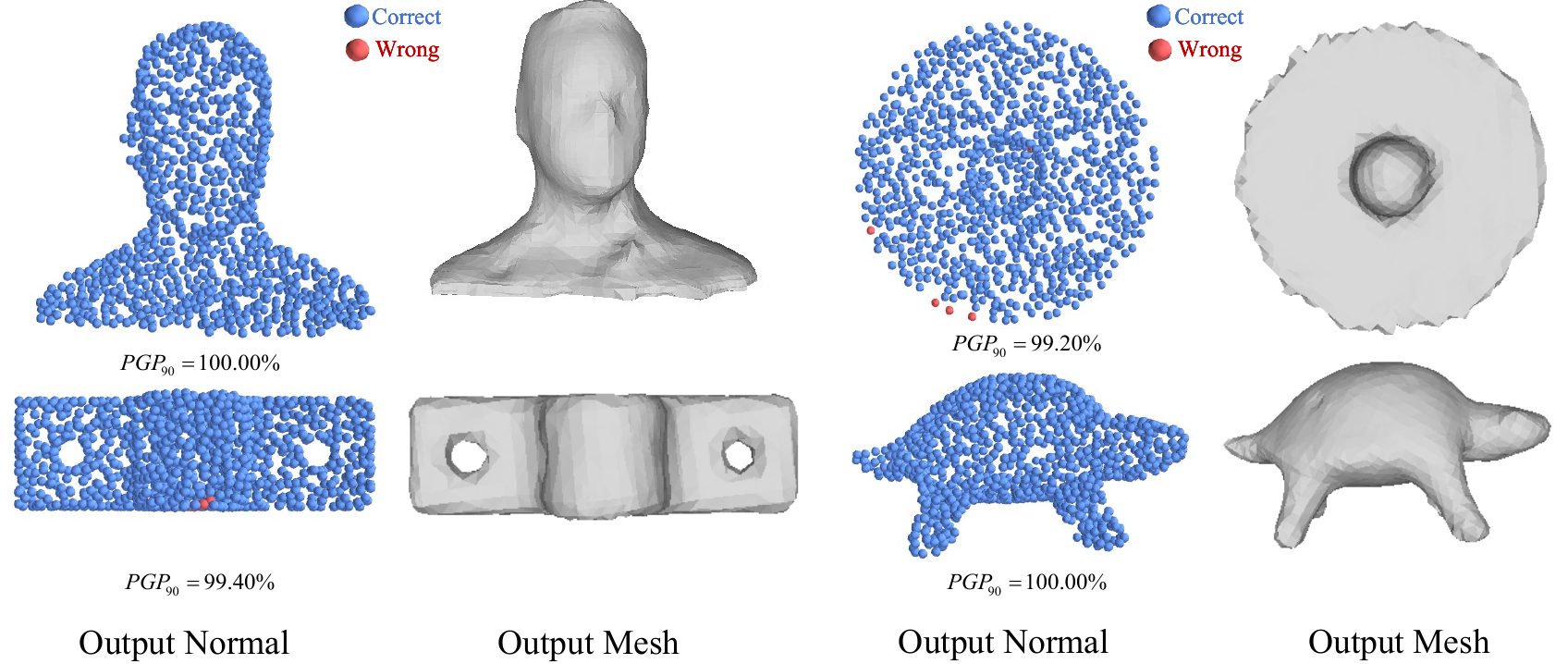}
    \caption{Qualitative and quantitative orientations of our method with the output meshes. The red points indicate the wrong orientation. Our method demonstrates good performances on orientation. }
    \label{orientation}
\end{figure}
\paragraph{\textbf{$N_{h} \geqslant 2 M$}}
At this point, the equation system becomes determined or even overdetermined. Hence, we seek the least squares solution of it. That is

\begin{align}\label{LSS}
\mathop{ \min }\limits_{\bm{\mu}} ||B\bm{\mu}-\bm{b}||_{2}^{2}.
\end{align}  

To avoid the numerical instability mentioned above and to alleviate the possible influence of noise, we
apply regularization to optimization problem \eqref{LSS} and to be
\begin{align}\label{LSS2}
\mathop{ \min }\limits_{\bm{\mu}} ||B\bm{\mu}-\bm{b}||_{2}^{2}+||\Gamma \bm{\mu}||_{2}^{2}.
\end{align}   

\begin{theorem}[\textbf{least-squares solution}]\label{TH4}
	The solution for  optimization problem  \eqref{LSS} is $\bm{\mu}=(B^TB+\Gamma^T\Gamma)^{-1}B^T\bm{b}$.
\end{theorem}
  \begin{proof}
  Let $f(\mu)=||B\bm{\mu}-\bm{b}||_{2}^{2}+||\Gamma \bm{\mu}||_{2}^{2}$, then $\nabla f=0$:
  \begin{align*}
      &B^{T}(B\bm{\mu}-\bm{b})+\Gamma^T\Gamma\bm{\mu}=0\\
      &(B^TB+\Gamma^T\Gamma)\bm{\mu}=B^T\bm{b}.
  \end{align*}
  
  Hence,
  \begin{align*}
      \bm{\mu}=(B^TB+\Gamma^T\Gamma)^{-1}B^T\bm{b}.
  \end{align*}
  
  On the other hand, $\forall \bm{y} \in \mathbb{R}^{3M}$,
  \begin{align*}
      &||B(\bm{\mu}+\bm{y})-\bm{b}||_{2}^2+||\Gamma(\bm{\mu}+\bm{y})||_{2}^2\\
      &=||B\bm{\mu}-\bm{b}||_{2}^{2}+2\bm{y}^TB^T(B\bm{\mu}-\bm{b})+||B\bm{y}||_{2}^{2}+||\Gamma \bm{\mu}||_{2}^{2}+||\Gamma \bm{y}||_{2}^{2}+2\bm{y}^T\Gamma^T\Gamma\bm{\mu}\\
      &=||B\bm{\mu}-\bm{b}||_{2}^{2}+||\Gamma \bm{\mu}||_{2}^{2}+||B\bm{y}||_{2}^{2}+||\Gamma \bm{y}||_{2}^{2}+2\bm{y}^T(B^TB\bm{\mu}+\Gamma^T\Gamma\bm{\mu}-B^T\bm{b})\\
      &=||B\bm{\mu}-\bm{b}||_{2}^{2}+||\Gamma \bm{\mu}||_{2}^{2}+||B\bm{y}||_{2}^{2}+||\Gamma \bm{y}||_{2}^{2}\\
      &\geqslant ||B\bm{\mu}-\bm{b}||_{2}^{2}+||\Gamma \bm{\mu}||_{2}^{2}.
  \end{align*}
  \end{proof}  
  
Given that the coefficient matrix is symmetric, we can employ the conjugate gradient method for its solution instead of calculating the inverse matrix. Like PGR, we use $N_{h}=2M$ to solve the problem and  the formulation \eqref{LSS2} is equal to the formulation \eqref{min3} in this situation.
Uniform regularization does not work well with nonuniform sampling. We discover choosing $\Gamma^T\Gamma$ as a multiple of the diagonal of $B^T B$ adapts to varying sampling density
\begin{align*}
    \Gamma^T\Gamma=\alpha\cdot 10M \cdot \diag(B^TB),
    \end{align*}
where $\alpha$ is a hyperparameter and $M$ is the number of input point clouds.    
Adding such regularization can lead to the better orientations and reconstructions, which is shown in \cref{5.2}.

\begin{figure}[htbp]
    \centering
    \includegraphics[width=1.0\linewidth]{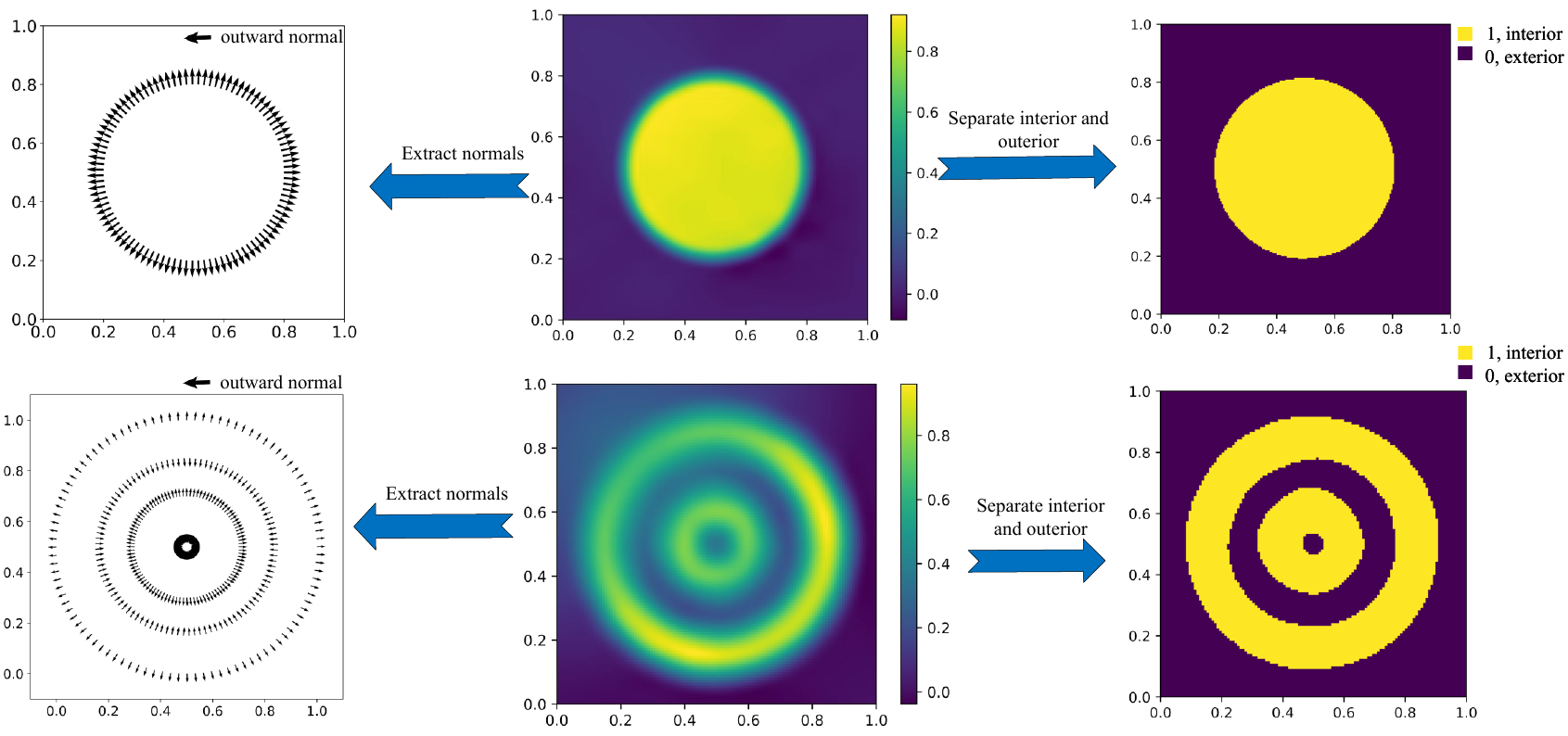}
    \caption{The qualitative result of our method to representing the indicator function and estimating the oriented normals in the two-dimensional case. In mathematical terms, the wavelet function's multiscale and compact support make our method work well for non-simply-connected regions, which also allows us to deal with
    complex topologies both in two-dimensional or three-dimensional cases.}
    \label{2D}
\end{figure}
\subsection{Normal and Orientation}
We reshape $\bm{\mu}$ into the matrix with the size of $M \times 3$, where each row represents an input point's normal information. That is
\begin{align*}
    \bm{\mu}\in \mathbb{R}^{3N\times 1} \to \widehat{\bm{\mu}}=\begin{bmatrix}
        \mu_{1}&\mu_{2}&\mu_{3}\\
        \vdots&\vdots&\vdots\\
        \mu_{3N-2}&\mu_{3N-1}&\mu_{3N}\\
    \end{bmatrix}\in \mathbb{R}^{N\times 3},
\end{align*}and let 
$(s_{\bm{p}_i},\widehat{\bm{n}}_{\bm{p}_{i}})$ be the estimated area element and normal for point $\bm{p}_{i} \in \mathcal{P}$,
\begin{equation}\label{NE}
\begin{aligned}
&s_{\bm{p}_i}=\sqrt{\mu_{3i+1}^{2}+\mu_{3i+2}^{2}+\mu_{3i+3}^{2}},
\\
&\widehat{\bm{n}}_{\bm{p}_{i}}=\dfrac{(\mu_{3i+1},\mu_{3i+2},\mu_{3i+3})}{s_{\bm{p}_i}}.
\end{aligned}
\end{equation}

Thus, we obtained a set of normals with consistent orientation $\{\widehat{\bm{n}}_{\bm{p}_{i}}\}_{i=1}^{M}$. Tables  \ref{1K} and \ref{10K} present the quantitative comparison of normal estimation with other famous methods. Our method demonstrates superior performance, showcasing the high-quality normals with minimal angular deviation from the ground truth. The results outperform other famous methods.

\subsection{Iso-surfacing}
After completing the normal estimation task, we calculate the value of the indicator function on the query points by averaging the results of the matrix multiplications 
 to reconstruct the surface. We select the set of all corner points on the octree $\mathcal{Q}=\{\bm{q}_{s}\}_{s=1}^{N_{Q}}$ as the query point set. That is
\begin{equation*}
\mathcal{Q}.\text{value}=B_{f}(\mathcal{Q};\mathcal{P})A(\mathcal{P};\mathcal{P})\bm{\mu}.
\end{equation*}

Note that the numerical calculation error may cause the slight shift of iso-value from $\frac{1}{2}$. We update the iso-value $v_{\text{iso}}$ with 
\begin{align*}
    v_{\text{iso}}
    &=\text{average}\left(B_{f}(\mathcal{P};\mathcal{P})\bm{c}_{f}\right)\\
    &=\text{average}\left(B_{f}(\mathcal{P};\mathcal{P})A(\mathcal{P};\mathcal{P})\bm{\mu}\right),
\end{align*}
where $\bm{c}_{f}$ are matrix of the corresponding
coefficients $c_i$ of the wavelet bases.

Finally, we extract surfaces with marching cubes by $v_{\text{iso}}$ and $\mathcal{Q}.\text{value}$. \textbf{A detailed description of our method is provided in Algorithm \ref{algorithm2}.} Our method demonstrates the state-of-the-art performance of
reconstruction, as shown in Figure \ref{5K2}, Tables  \ref{1K} and \ref{10K}. 

\begin{algorithm}[htbp]
\caption{Wavelet-based Global Orientation and Surface Reconstruction}
\label{algorithm2}
\textbf{Input:} Unoriented point cloud $\mathcal{P}$, wavelet function's maximum depth $D_{\max}$, smooth width $\varepsilon$, the number of homogeneous constraint $N_{h}$, regularization factor $\alpha$. \\
\textbf{Output:} Oriented point cloud $(\mathcal{P},\mathcal{N})$, a watertight surface $M$ approximating the points.
\begin{enumerate}
\item Initialize $\bm{\mu}=\bm{0}$, $\bm{b}_1=\frac{\mathbbm{1}}{2}$.
\item \textit{// Mollify the wavelet basis functions}
\item For $j = 1$ to $D_{\max}$:
\begin{enumerate}
\item Calculate the mollifying kernel function $K_{2^j \varepsilon}(t)$;
\item Calculate $\overline{\psi}_{\varepsilon,j}=2^{\frac{j}{2}}\overline{\psi}_{2^j \varepsilon}(2^j x)$;
\end{enumerate}
\item \textit{// The homogeneous and non-homogeneous constraint}
\item Construct the homogeneous constraints $H\bm{\mu}=\bm{b}_2$, $\bm{b}_2=\bm{0}$ by Equation \eqref{BA};
\item Construct the non-homogeneous constraints $B_{f}A \bm{\mu}=\bm{b}_1$;
\item Let $B=\begin{bmatrix} B_{f}A \\ H \end{bmatrix}$, $\bm{b}=\begin{bmatrix} \bm{b}_1 \\ \bm{b}_2 \end{bmatrix}$;
\item \textit{// Conjugate gradient (CG) for the LSE}
\item If $N_{h}\leqslant 2|\mathcal{P}|$:
\begin{enumerate}
\item Calculate $B_{c}=BB^{T}+\text{diag}(BB^T)$;
\item Solve $B_{c}\bm{\xi}=\bm{b}$ for $\bm{\xi}$ using conjugate gradient method;
\item Set $\bm{\mu}= B^T \bm{\xi}$
\end{enumerate}
\item Else:
\begin{enumerate}
\item Calculate $B_{c}=B^TB+\text{diag}(B^TB)$;
\item Calculate $\bm{d}=B^T\bm{b}$;
\item Solve $B\bm{\mu}=\bm{d}$ for $\bm{\mu}$ using conjugate gradient method;
\end{enumerate}
\item \textit{// Orientation and surface reconstruction}
\item Calculate normals $\bm{n}_{i}=\dfrac{\bm{\mu}_i}{||\bm{\mu}_i||_{2}}$, and output oriented points $(\mathcal{P},\mathcal{N})$;
\item Calculate the iso-value as $v_{\text{iso}} = \text{average}\left(A(\mathcal{P}; \mathcal{P})\bm{\mu}\right)$;
\item Obtain the values of query points as $\mathcal{Q}.\text{value} =A(\mathcal{Q}; \mathcal{P})\bm{\mu}$;
\item Reconstruct surface $M$ using marching cubes by $\mathcal{Q}.\text{value}$ and $v_{\text{iso}}$.
\end{enumerate}
\end{algorithm}

\section{Experimental Results}
\label{sec:experiments}

\begin{table*}\small
\centering
\caption{Quantitative comparisons of our method with other state-of-the-art methods on the orientation and reconstruction in sparse point clouds. The Chamfer Distance (CD) values are multiplied by $10^{4}$. The best results are in bold. Our method outperforms the baselines in the statistical sense.}
\label{table2}

  \centering
    \begin{tabular}{|cl|cccc|cc|}
    \hline
    \multicolumn{2}{|c}{}&\multicolumn{4}{|c|}{1K,noise:0\%}&\multicolumn{2}{c|}{1K,noise:0.5\%}\\
        \multicolumn{2}{|c|}{} & Realworld  & Famous & ABC   & Thingi10k   & Realworld  & Famous  \\
    \hline
    \multirow{7}{*}{$\text{PGP}_{90}$ $\uparrow $} 
    &PCA  + MST \cite{hoppe1992surface}&0.7613&0.7589&0.7221&0.7599&0.7279&0.7404\\
    &Dipole \cite{dipole} & 0.8000 & 0.8469 & 0.7354 & 0.8580  &0.7923 &0.8338 \\
    &iWSR \cite{iwsr}&0.8737 & 0.8711& 0.8167& 0.9070&0.8690&0.8677\\
    & GCNO \cite{2023GCNO} &   0.9433    &   \textbf{0.9211}    &  0.8642     & 0.9436 &0.9364    &   \textbf{0.9169}    \\
       & iPSR \cite{hou2022iterative} & 0.9390 & 0.9084 & 0.8513 & 0.9344  &0.8598 & 0.8616  \\
         & Ours  & \textbf{0.9440} & 0.9087 &\textbf{0.8662}&\textbf{0.9437}&\textbf{0.9367}&0.8951 \\
    \hline
       
    \multirow{7}{*}{CD $\downarrow $} 
    &PCA + MST \cite{hoppe1992surface}&197.33&87.83&524.15&214.12&233.98&84.64\\
    &Dipole \cite{dipole}&5.55&11.52&93.84&32.46&8.55&19.52\\
    &iWSR  \cite{iwsr}&46.80&57.58&218.45&78.49&59.25&29.77\\
    & GCNO \cite{2023GCNO} &   5.33     &  \textbf{4.33}    &     13.56    &  9.61  &  5.89     &   \textbf{7.36} \\
         & iPSR \cite{hou2022iterative}  & 10.96 & 5.71  & 13.25  & 9.27 & 20.37 & 12.59\\
          & Ours  & \textbf{4.95} & 7.13 & \textbf{12.96} & \textbf{7.02} &\textbf{5.74}&10.62\\
    \hline
    \end{tabular}%
\label{1K}
\end{table*}


\paragraph{\textbf{Experimental Setup}}  Algorithm \ref{algorithm2} shows the complete pipeline of our method. The equation solving stage mainly involves matrix
multiplications, for which we use Numpy \cite{harris2020array} for auxiliary processing on the CPU.
Experiments are conducted on  a desktop PC with Intel i7-10870H CPU. 

\paragraph{\textbf{Evaluating Indicator}}
		The proportion of good points (PGP)  is the ratio of points for which the angle between the estimated normal and the ground truth normal is smaller than a specified threshold. A higher value of PGP indicates the better accuracy of the orientation.
		\begin{equation*}
        \begin{aligned}
			&\text{PGP}_{90}(P)=|\text{correct}.P|/|P|, \\
            &\text{correct}.P=\{\bm{p}_{i} \in P | \  \bm{n}_{\bm{p}_{i,\text{out}}} \cdot \bm{n}_{\bm{p}_{i,\text{true}}}>0\}.
        \end{aligned}
		\end{equation*}
		
		The Chamfer distance (CD)  penalizes both false negatives (missing parts) and false positives (excess parts) to evaluate the reconstruction error,
		\begin{equation*}
			\text{CD}(S_{1},S_{2})=\dfrac{1}{|S_{1}|}\sum_{\bm{x} \in S_{1}}\mathop{\min }\limits_{\bm{y} \in S_{2}}||\bm{x}-\bm{y}||_{2}^{2}+\dfrac{1}{|S_{2}|}\sum_{\bm{y} \in S_{2}}\mathop{\min }\limits_{\bm{x} \in S_{1}}||\bm{x}-\bm{y}||_{2}^{2}.
		\end{equation*}
  
    $S_{1}$ and $S_{2}$ denote the reconstructed and ground truth surfaces, respectively. $|\cdot|$ represents the number of elements. To evaluate the quality of the reconstruction, we adopt 20K sample points for either surface.

    \paragraph{\textbf{Parameters}}
We adopt the parameter setting  $\alpha=2$ for clean point clouds, $\alpha=2.5 $ for point clouds with 0.5\% Gaussian noise. We use the CG algorithm to solve the linear system.

\paragraph{\textbf{Baselines}} We include methods (PCA+MST \cite{hoppe1992surface}, Dipole \cite{dipole}, iWSR \cite{iwsr}, GCNO \cite{2023GCNO}, iPSR \cite{hou2022iterative}) for comparison. 
For PCA, Dipole and GCNO, we follow the default setting and match it with SPR \cite{kazhdan2013screened} for reconstruction. For iWSR, despite selecting numerous parameters, its efficiency significantly decreases when dealing with sparse point cloud datasets. For iPSR, we have tried our best to find their optimal parameters and show the optimal results in this paper.

\subsection{Experimental Effect}
\paragraph{\textbf{2D Cases}}
Our method is based on the divergence theorem and constructing the wavelet basis functions by tensor product. In addition to three-dimensional cases, our method can also represent the indicator function of the closed area in two-dimensional cases and estimate normals for points on the boundary. Moreover, two-dimensional case can show the representing quality of our method on the indicator function more intuitively. We select different shapes such as spheres and multi-layer circular rings for experiments, and the results are shown in Figure \ref{2D}.

	\begin{table*}[tbph] \scriptsize
			\centering
			\resizebox{0.999\textwidth}{!}{
			\begin{tabular}{|l|cccccc|cccccc|}
				\hline
				\multirow{2}{*}{Models}& \multicolumn{6}{c|}{$\text{PGP}_{90}$  $\uparrow $} & \multicolumn{6}{c|}{ CD  $\downarrow$ } \\

				& Ours   & iPSR & GCNO & iWSR & Dipole & PCA & Ours   & iPSR & GCNO & iWSR & Dipole & PCA \\
 \hline

    10K81368 & \textbf{1.000} & 0.992 & \textbf{1.000} & 0.999 & 0.944 & \textbf{1.000} & \textbf{1.64} & 2.13 & \textbf{1.64}  & 3.01  & 5.07  & 1.70 \\
    10K75653 & \textbf{1.000} & 0.999 & \textbf{1.000} & 0.949 & 0.992 & 0.796 & \textbf{1.62} & 1.66 & 2.89  & 4.15  & 8.95  & 19.1 \\
    10K70558 & \textbf{1.000} & \textbf{1.000} & \textbf{1.000} & 0.999 & 0.998 &  0.999 & \textbf{3.07} & 3.19 & 3.26 & 6.10 & 6.22 & 6.05\\
    10K47984 & \textbf{1.000} & 0.997 & 0.999 & 0.533 & 0.999  & \textbf{1.000}  & \textbf{0.94}  & 0.97 & 0.99 & 1.87& 1.40 &1.93\\
    10K75496 & 0.998 & 0.998 & \textbf{1.000} & 0.996 & 0.998 & 0.997 & 1.04 & 1.09 & \textbf{1.01} & 1.74 & 1.77 &1.65\\
    10K98480 & \textbf{0.994} & \textbf{0.994} & 0.993 & 0.958 & 0.986 & 0.864 & \textbf{1.88}  & 1.89  &  1.92 &7.29 & 4.88 &15.00\\
    Tortuga & \textbf{1.000} & 0.998 & 0.546 & \textbf{1.000} & 0.999 & 0.999 & \textbf{0.77} & 0.81  & 12.66  & 1.52  & 1.47 &1.58\\
    Armadillo & \textbf{0.988} & 0.978 & 0.986 & 0.947 & 0.987 & 0.967 & \textbf{1.42}  & 1.46  & 1.87 & 8.89 & 1.53 & 3.62\\
    ABC17633 & \textbf{1.000} & \textbf{1.000} & 0.996 & 0.499 &  0.701 &  0.496 & \textbf{0.72}  & 0.74 & 0.88 & 52.65 & 14.95 & 27.99\\
    ABC14952 & \textbf{1.000} & 0.999 & \textbf{1.000} & 0.972 & 0.999 & \textbf{1.000} & 1.71 & 1.72 & 1.71 & 3.69  & 1.99 &\textbf{1.67}\\
    ABC993520 & \textbf{1.000} & \textbf{1.000} & \textbf{1.000} & \textbf{1.000} & 0.997 & \textbf{1.000}  & 0.74  & \textbf{0.68}  & 0.69 &1.61 & 1.55 &1.69 \\
    ABC19536 & 0.998 & 0.995 & 0.995 & 0.956 &  \textbf{1.000} & \textbf{1.000} & 1.15 & \textbf{1.00} & 1.18 & 1.45 & 3.65 & 1.70\\
    ABC11602& \textbf{0.996} & 0.874 &  \textbf{0.996} & 0.851 &  0.945 & \textbf{0.996} & \textbf{2.08}  & 6.18  & 2.24  & 8.25 & 7.93 & \textbf{2.08} \\
       \hline
    Average & \textbf{0.998} & 0.988 & 0.969 & 0.939 & 0.967  & 0.937 & \textbf{1.55} & 2.16  & 2.36 & 7.60 & 4.38 &6.67 \\
 \hline
			\end{tabular}%
            }
			\caption{Comparison of our method with other state-of-the-art methods for point clouds with 5K points on orientation and reconstruction. The CD values are multiplied by $10^{4}$. Our method exhibits superior performance against other state-of-the-art methods.}
			\label{10K}
		\end{table*}%
The output mollified indicator function of our method  is shown in the middle subgraph. Our method can separate out the interior and exterior of regions distinctly and effectively, which is shown in the right subgraph. Even when dealing with intricate and disconnected regions like multilayered rings, our method can output the results with remarkable quality consistently. Our method can also extract the uniformly oriented normals from $\bm{\mu}$, which is shown on the left subgraph.

\begin{figure}[htbp]
    \centering
    \includegraphics[width=0.7\linewidth]{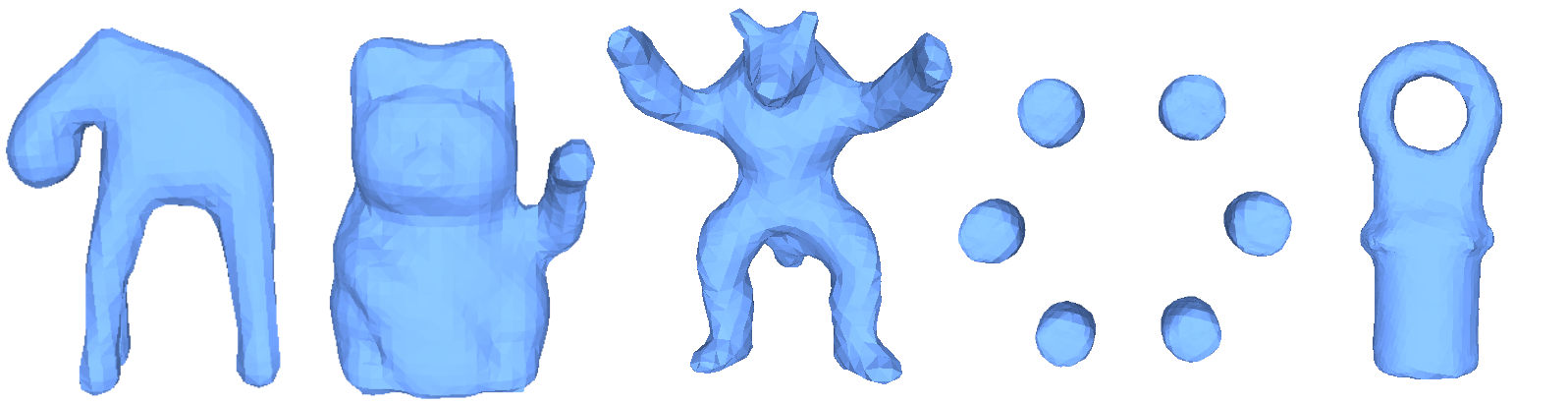}
    \caption{Qualitative result of our method on the reconstruction of sparse models with 1K points. Our
    method can deal with reconstruct surfaces with good quality and compensate for the shortcomings of other state-of-the-art surface reconstruction methods based on wavelet basis functions.}
    \label{1k2}
\end{figure}

\paragraph{\textbf{Famous, ABC and Thingi10K Datasets}}
The Famous \cite{erler2020points2surf} dataset includes a variety of classic shapes such as bunny, dragon, and armadillo. The ABC \cite{kingma2014adam} dataset comprises a diverse collection of CAD meshes, while the Thingi10K \cite{zhou2022learning} dataset contains a variety of shapes with complex geometric details. We randomly sample 1K points from each model to test our method capabilities on sparse point clouds. The quantitative comparisons of all methods are shown in Table \ref{1K}.  The qualitative results of our method are shown in Figure \ref{1k2}. Our method demonstrates better performance in normal estimation and reconstruction, particularly
in sparse models.

\begin{figure}[htbp]
    \centering
    \includegraphics[width=1\linewidth]{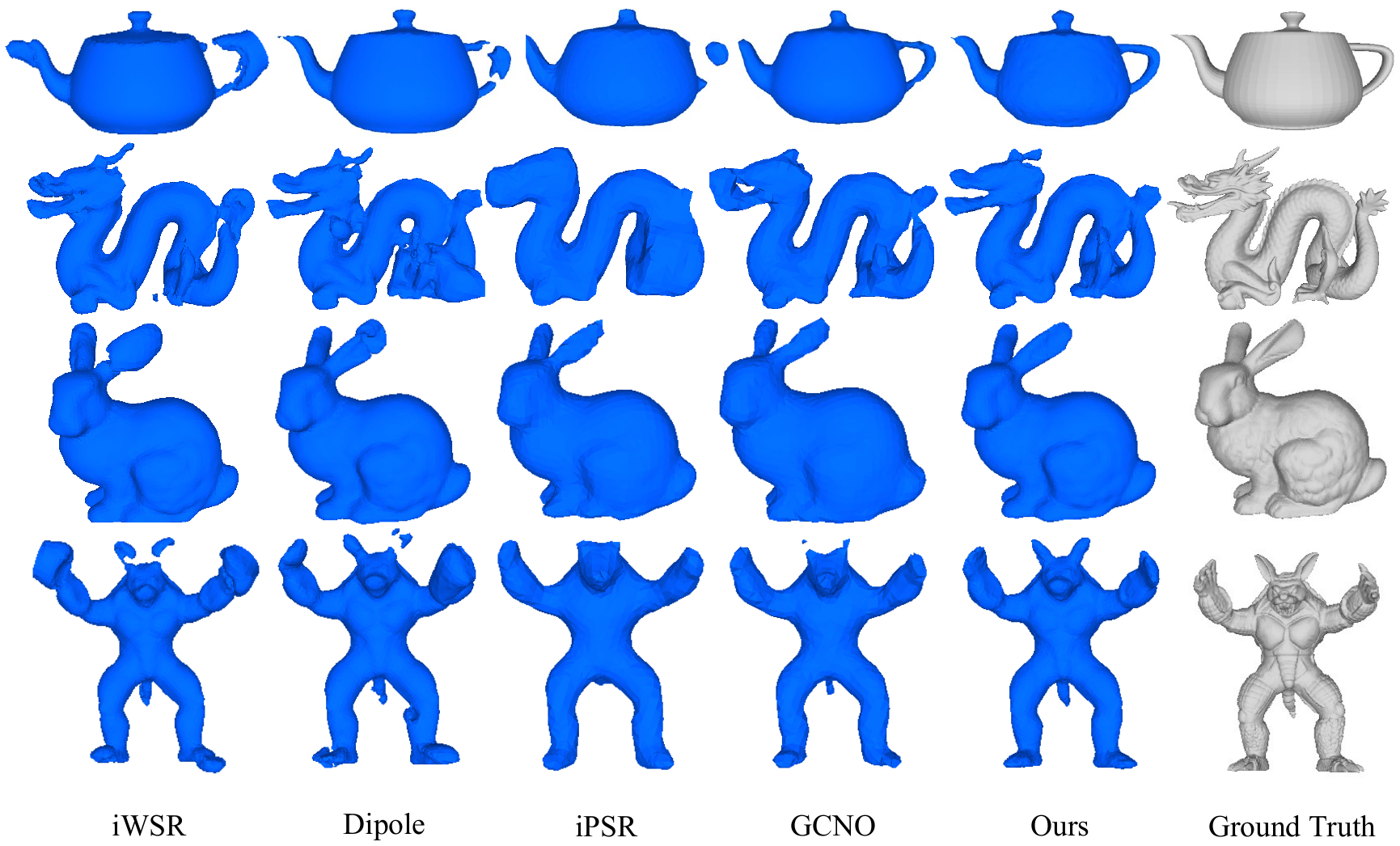}
    \caption{ Qualitative comparison of our method and baselines on
    reconstruction for models. Our method outperforms the
    baselines.}
    \label{5K2}
\end{figure}

We also evaluate our method with dense point clouds to demonstrate its ability to reconstruct complex structural details accurately. Table \ref{10K} shows some results for models with 5K points. The qualitative comparison of surface reconstruction on these datasets is shown in Figure \ref{5K2}. In addition, the qualitative comparison of normal estimation and orientation on these datasets is also provided in Figure \ref{orientation}.
Experimental results demonstrate the state-of-the-art performance of our method on these datasets.

\paragraph{\textbf{Real-world Dataset}}
The models in Real-world \cite{huang2019variational} dataset contains noise and outliers. In addition, the ground truth mesh of such models exhibits lower smoothness. To verify the robustness of our method, we utilize the Real-world dataset to generate point clouds with 1K  points. The quantitative comparison results of the methods are shown in Table \ref{1K}. Our method outperforms the baselines in the average sense.

\begin{figure}[htbp]
    \centering
    \includegraphics[width=1\linewidth]{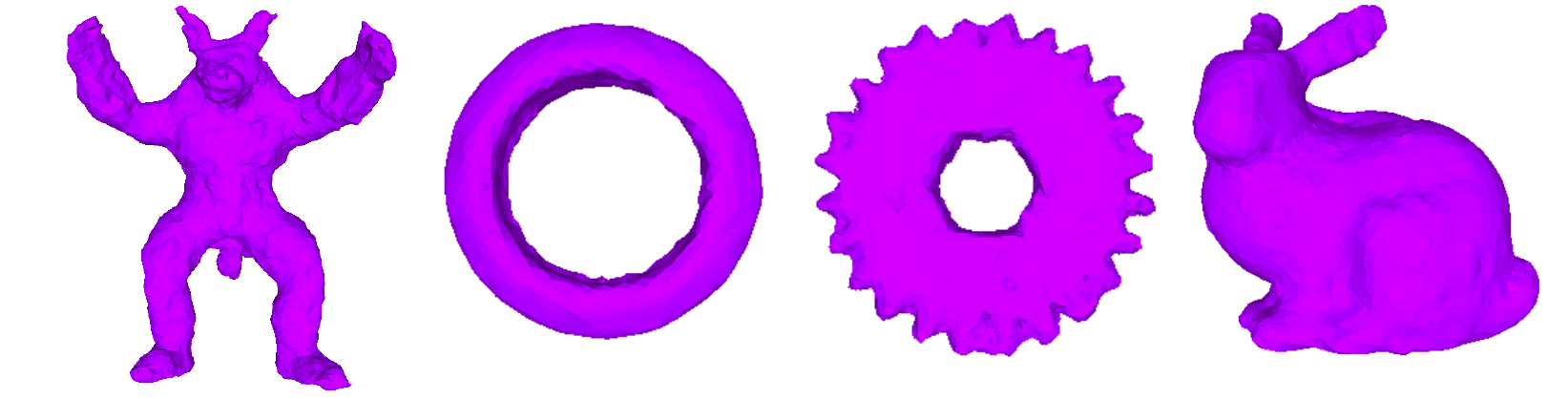}
    \caption{Qualitative result of our method on the reconstruction of noisy models with 2K points. Our method can effectively handle noisy data and output high-quality surfaces.}
    \label{noisy}
\end{figure}

\paragraph{\textbf{Noisy Datasets}}
Evaluating performance on noisy datasets is crucial for demonstrating the robustness of algorithms against disturbances. However, most non-learning methods cannot handle noisy point clouds well, especially in sparse point clouds. We introduce a uniform Gaussian noise of 0.5\% to the Famous \cite{erler2020points2surf}, ABC \cite{kingma2014adam}, Thingi10K \cite{zhou2022learning}, and Real-world \cite{huang2019variational} datasets with 1K points as input. The quantitative comparisons of our method and other well-known methods are shown in Table \ref{1K}. Figure \ref{noisy} displays the reconstructions from noisy point clouds. Our method outputs globally consistent orientations and achieves good performance in denoising and preserving detail.
\paragraph{\textbf{Wireframe Models}}

\begin{figure}[htbp]
    \centering
    \includegraphics[width=1.0\linewidth]{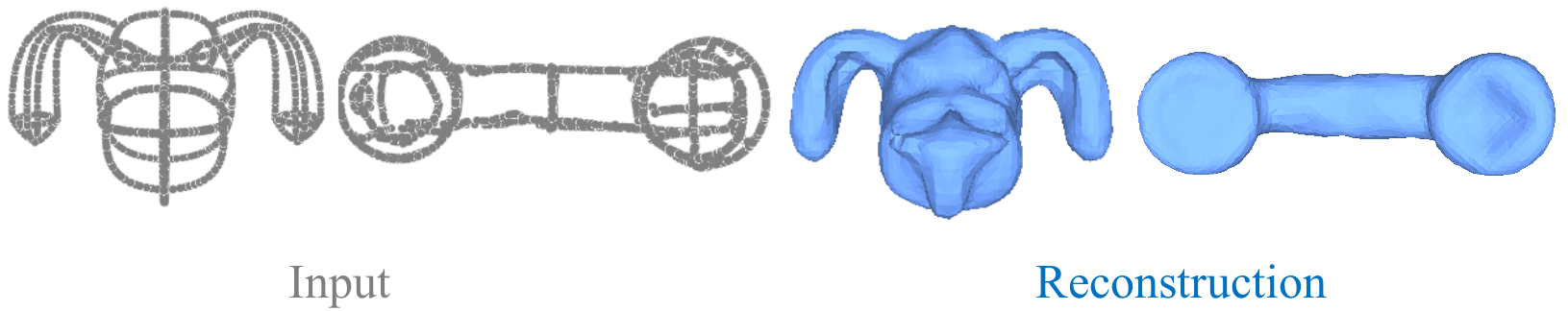}
    \caption{Our method's reconstruction and orientation results of
wireframe models. All the models are from VIPSS \cite{huang2019variational}.  The input unoriented point clouds (1K) are shown in the left of figure. The 
reconstructed surfaces generated by our method are shown in the right of figure.}
    \label{m}
\end{figure}

Sparse and wireframe points can be another challenge due to the lack of detailed information. Wireframe models are more difficult to reconstruct due to their sparse point cloud (1K) and regular arrangement, which results missing points in most areas. Even slight deviations in normals can result in severe reconstruction errors. Figure \ref{m} illustrates the good reconstruction results of our method  for these models.

\begin{figure}[htbp]
    \centering
    \includegraphics[width=0.7\linewidth]{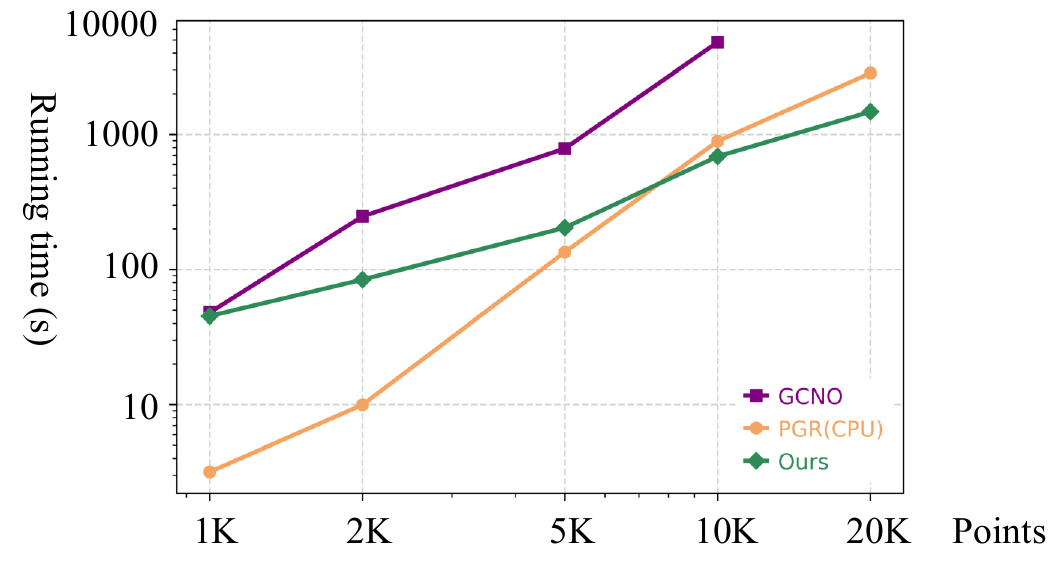}
    \caption{Qualitative comparison of the running time of our method and the baselines. GCNO takes hours to converge on a 20K point cloud. Benefited from the  compactness and orthogonality of wavelet basis functions, our method controls the computation time in CPU within an acceptable range. }
    \label{times}
\end{figure}

\paragraph{\textbf{Running Speed}}
The running speed  of algorithms is a crucial indicator. Benefited from the orthogonality and compact support of the Db4 wavelet basis function, the computational complexity of our method in establishing equations is greatly reduced, which enables our method to handle the orientation and reconstruction tasks for unoriented point clouds on a laptop CPU. The quantitative comparison of running speed is shown in Table \ref{times}.

 GCNO struggles to handle dense point clouds and takes hours to converge on a 20K point cloud. Our method controls the computation time in CPU within an acceptable range. In addition, our method has the potential to achieve further acceleration by associating wavelet basis functions with octree structures.

\subsection{Evaluation of Individual Parameters}
\label{5.2}
\paragraph{\textbf{Homogeneous Constraint}}
By using the divergence-free function fields, we construct homogeneous constraints on linearized surface elements $\bm{\mu}$ to introduce more surface information. In Table \ref{3}, we provide a detailed presentation of the quality changes in orientation resulting from the addition of the homogeneous constraints. 

The quality of orientation  significantly improves by adding the homogeneous constraint. By introducing $2M$ additional equations, sparse point clouds can also have achieved high-quality orientation. However, we also observe that adding constraints can lead to marginal effects. Furthermore, introducing an excessive number of constraints can actually decrease the overall quality of the equations due to numerical errors and the similarity of constraints, consequently reducing the final orientation accuracy.

\begin{table}[htbp]
  \centering
  \caption{The quantitative comparison of our method's orientation with different quantity $N_{h}$ of homogeneous constraint.}
    \begin{tabular}{l|cccccc}
    \hline
    $\text{PGP}_{90}$ $\uparrow $     & 0     & $M$   & $1.5M$     & $2M$     & $2.5M$    & $3M$ \\
        \hline
    Bunny & 0.8460  & 0.9675 &  0.9770 & \textbf{0.9785}  & 0.9715 &  0.9700 \\
    Dragon & 0.8060 & 0.9050 & 0.9060 &  \textbf{0.9170}  & 0.9090 & 0.9070  \\
    Utah\_teapot & 0.9220 & 0.9385 & 0.9455 & \textbf{0.9600} & 0.9560 &  0.9460 \\
    Armadillo & 0.8700 & 0.9560 & 0.9680 & \textbf{0.9720} & 0.9685 &  0.9670 \\
    
    \hline 
    \end{tabular}%
  \label{3}%
\end{table}%
\paragraph{\textbf{Regularization Factor}}
Although we have mollified the wavelet basis functions, it is still difficult to completely eliminate the ambiguity in the final linear equations. The regularization factor plays a crucial role in enhancing the quality of the equations. Table \ref{4} shows the reconstructions of sparse models with different $\alpha$. On one hand, without regularization, artifacts appear even for such clean models and the orientation quality falls well below the expected level. On
the other hand, if regularization factor is too large, the regularized system of equations deviates  from the initial system of equations significantly, causing the solution $\bm{\mu}$ to lose the geometric interpretation of the normal multiplied by the area element. The orientation and reconstruction ability of our method is also reduced.
As shown in Equation \eqref{LSS2}, the addition of the regularization term is equivalent to introducing an additional $L_2$-norm loss to the existing constraints. This implies that as the regularization term $\alpha$
 increases, the $L_2$-norm of the solved $\bm{\mu}$ tends to minimize, resulting in a smoother surface. However, excessively large values of the regularization term $\alpha$
 can lead to the loss of surface details, thereby reducing the quality of the reconstruction.
In general, as $\alpha$ increases, the
reconstruction first improves and then degrades in quality.

\begin{table}[htbp]
  \centering
  \caption{The quantitative comparison of our method's orientation under different regularization factor $\alpha$.}
    \begin{tabular}{l|ccccccc}
    \hline
    $\text{PGP}_{90}$ $\uparrow $     & 0     & 0.5   & 1     & 1.5     & 2    & 3 & 4\\
        \hline
    Bunny & 0.6985 & 0.9635 & 0.9730 & \textbf{0.9795} & 0.9785 & 0.9735 & 0.9725\\
    Dragon & 0.5545 & 0.8890 & 0.9010 & 0.9160 & \textbf{0.9170} & 0.9040 & 0.9015 \\
    Utah\_teapot & 0.5785 & 0.9375 & 0.9380 & 0.9575 & \textbf{0.9600} & 0.9590 & 0.9585\\
    Armadillo & 0.6680 & 0.9610 & 0.9660 & 0.9660 & \textbf{0.9720} &  0.9670 & 0.9585 \\
    
    \hline 
    \end{tabular}%
  \label{4}%
\end{table}%




\section{Limitations and Conclusion}
\label{sec:conclusions}
Based on wavelet basis functions with compact support and orthogonality, we proposes a novel strategy for representing the mollified indicator function. Furthermore, through rigorous mathematical derivations, we transfer the modifying calculation to wavelet basis functions, improving the accuracy of coefficient computation while maintaining the multi-resolution of wavelet basis functions. This ultimately enables high-quality wavelet surface reconstruction and orientation.

However, as a method based on indicator function, our method necessitates the target area $\Omega$ to be a bounded region, implying that the surface $\partial \Omega$ must be closed. Similar to other implicit methods, our method is unable to accommodate 3D shapes that do not conform to this condition, such as non-manifold surfaces and open scans. In addition, although we have designed a dedicated acceleration algorithm based on the Db4 wavelet basis functions, a possible future direction is to implement a more advanced acceleration method using GPU and the fast multipole method (FMM), which can further reduce complexity. However, the implementation of FMM requires more meticulous mathematical analysis and deals with challenges in terms of implementation and parallelization, especially concerning the  multi-resolution of wavelet. The substantial amount of work involved is sufficient to consider it as a separate future work.

Overall, our method has significantly improved the effectiveness of wavelet-based surface reconstruction and orientation at an acceptable speed, particularly for sparse point cloud models lacking sufficient information to represent the indicator function. Paired with iWSR, we have developed a comprehensive framework based on wavelet analysis for the reconstruction and orientation of unoriented point clouds. Such framework is designed to handle point clouds of any scale effectively.

\appendix
\section{The compact support and orthogonality of Db4 wavelet bases}
\label{Appendix}
\begin{theorem}[\textbf{Construction orthonormal wavelet sequences}]\label{th1}
	Let $\{V_{j}\}$ be an $MRA$ with scaling function $\varphi(x)$ and scaling filter $h(k)$, where $V_{j}=\{f(x):f(x)=2^{\frac{j}{2}}\varphi_{0}(2^{j}x-k),\varphi_{0}(x)\in V_{0}\}$, $V_{0}=\overline{\text{span}}\{\varphi(x-n)\},n \in \mathbb{Z}$.
	Define the wavelet filter $g(k)$ by $g(k)=(-1)^{k}\overline{h}(1-k)$, and the wavelet $\psi(x)$ by 
	$\psi(x)=\sum_{k}g(k)2^{1/2}\varphi(2x-k)$. Then $\{\psi_{j,k}(x)\}_{j,k \in \mathbb{Z}}$ is a set of one-dimensional orthogonal sequences, where $\psi_{j,k}(x)=2^{\frac{j}{2}}\psi(2^{j}x-k)$.  Alternatively, given any $J \in \mathbb{Z}$,$\{\varphi_{J,k}(x)\}_{k \in \mathbb{Z}} \bigcup \{\psi_{j,k}(x)\}_{k \in \mathbb{Z},j\geqslant J}$ is a set of one-dimensional orthogonal sequences, where $\varphi_{j,k}(x)=2^{\frac{j}{2}}\varphi(2^{j}x-k)$.
	\end{theorem}

	\begin{theorem}[\textbf{The support of the scaling function}]\label{th12}
	Suppose that $h(k)$ is a finite QMF, let $m_{0}(\gamma)$ be given by $m_{0}(\gamma)=\frac{1}{\sqrt{2}}\sum_{k}h(k)e^{-2\pi \mathrm{i} k \gamma}$, and holds that there is a number $c > 0$ such that $|m_{0}|(\gamma) \geqslant c $ for $|\gamma| \leqslant \frac{1}{4}$. Suppose that for some $N\in \mathbb{N}$, $h(k)$ has length $2N$. Then, the scaling function $\varphi(x)$ defined by the two-scale expansion equation is supported in an interval of length $2N- 1$.
	\end{theorem}



\bibliographystyle{siamplain}
\bibliography{sample-base}

\end{document}